\documentclass[xcolor=pdftex,usenames,dvipsnames,table]{PoS}

\usepackage{multirow, lscape}
\usepackage{multicol}
\usepackage{array}
\usepackage{graphicx}
\usepackage{amssymb,amsmath}
\usepackage{bm}
\usepackage{tcolorbox}

\usepackage{url}
\title{Nucleon Axial Form Factors from Clover Fermion on $2+1+1$-flavor HISQ Lattice}

\ShortTitle{Nucleon Axial Form Factors from Clover-on-HISQ}

\author{\speaker{Yong-Chull Jang}\\
        Physics Department,
        Brookhaven National Laboratory, 
        Upton, NY, 11973, U.S.A.\\
        E-mail: \email{ypj@bnl.gov}}

\author{Rajan Gupta, Tanmoy Bhattacharya, Sungwoo Park\\
        Theoretical Division, T-2, 
        Los Alamos National Laboratory, 
        Los Alamos, NM, 87545, U.S.A.\\
        E-mail: \email{rg@lanl.gov}}

\author{Boram Yoon\\
        Computer, Computational, and Statistical Sciences, CCS-7, 
        Los Alamos National Laboratory,
        Los Alamos, NM, 87545, U.S.A.}

\author{Huey-Wen Lin\\
        Department of Physics and Astronomy, 
        Michigan State University, MI, 48824, U.S.A}

\abstract{The nucleon axial form factors -- axial $G_A$, induced
  pseudoscalar $\widetilde{G}_P$ and pseudoscalar $G_P$ -- have
  displayed large systematics in lattice QCD calculations. The major
  symptoms were the violation of the partially conserved axial current
  (PCAC) relation between the three form factors, and the
  underestimation of the induced pseudoscalar coupling $g_P^\ast$ and
  the axial charge radius $r_A$ compared to phenomenological
  estimates. The small $g_P^\ast$ was a consequence of the failure of
  the pion-pole dominance (PPD) hypothesis, especially at low
  $M_\pi^2$. The small charge radius $r_A$ and the underestimate of
  $g_A$ were related. The dominant systematic responsible is the lack
  of inclusion of low-energy ($N \pi$) states that are not manifest in
  the multiexponential fit to the nucleon two-point correlator. We
  show that this low-energy state can be determined from the
  three-point correlator $\langle N A_4 N \rangle $ with the insertion of the temporal component
  of the axial current $A_4$ within the nucleon state, ie, the strategy labeled
  $S_{A4}$~\cite{Jang:2019vkm}. Including this low-energy state in
  fits to control excited-state contamination (ESC) gives results for
  $g_A$, $r_A$, and $g_P^\ast$ that are consistent with
  experimental/phenomenological values. However, the systematic
  uncertainties, especially in data at small $Q^2$, are now much
  larger.}

\FullConference{37th International Symposium on Lattice Field Theory - Lattice2019\\
		16-22 June 2019\\
		Wuhan, China}

\graphicspath{{./figs_arxiv/}}

\providecommand{\abs}[1]{\lvert#1\rvert}

\providecommand{\matrixe}[3]{\langle#1\lvert#2\rvert#3\rangle}
\providecommand{\expv}[1]{\langle#1\rangle}

\providecommand{\GeV}{\mathrm{GeV}}

\providecommand{\fm}{\mathrm{fm}}

\newcolumntype{C}[1]{>{\centering\arraybackslash}p{#1}}

\providecommand{\cut}[1]{{}}

\newcommand{\Dslash}{\ensuremath{{D\kern -0.65em /}}}

\providecommand{\MeV}{\mathrm{MeV}}

\definecolor{indigo}{HTML}{4B0082}
\definecolor{darkorange}{HTML}{FF8C00}
\definecolor{darkgreen}{HTML}{006400}
\definecolor{myred}{HTML}{ED3B3B}
\definecolor{mygreen}{HTML}{2E8B57}
\definecolor{myblue}{HTML}{2E74FF}
\definecolor{mybgc}{HTML}{FFFACD}

\begin{document}

\section{Introduction}
\label{sec:intro}

Nucleon matrix elements of the axial $A_\mu =
\bar{u}\gamma_\mu\gamma_5 d$ and pseudoscalar $P = \bar{u}\gamma_5 d$
currents can be decomposed into the axial $G_A$, induced pseudoscalar
$\widetilde{G}_P$, and pseudoscalar $G_P$ form factors as 
\begin{align}
  \matrixe{N(\vec{p}_f)}{A_\mu (\vec{Q})}{N(\vec{p}_i)} &=
  {\overline u}(\vec{p}_f)\left[ G_A(Q^2) \gamma_\mu
  + q_\mu \frac{\widetilde{G}_P(Q^2)}{2 M}\right] \gamma_5 u(\vec{p}_i) \,,
  \label{eq:aff-a}
  \\
  \matrixe{N(\vec{p}_f)}{P (\vec{q})}{N(\vec{p}_i)}  &=
  {\overline u}(\vec{p}_f)\left[ G_P(Q^2) \gamma_5 \right]u(\vec{p}_i) \,,
  \label{eq:aff-ps}
\end{align}
where $q = p_{f} - p_{i}$, and the initial nucleon is at rest
$\vec{p}_i=0$ in our lattice calculation. The space-like four momentum
transfer $Q^2 (= -q^2) = \vec{p}_f^2 - (E(\vec{p}_f)-M)^2$. Note that
we work in the isospin limit, $m_u = m_d$, and present results only
for the isovector currents, e.g.,
$A_\mu^{u-d}=\bar{u}\gamma_\mu\gamma_5 u - \bar{d} \gamma_\mu\gamma_5
d$ so that $\matrixe{p}{A_\mu}{n} = \matrixe{p}{A_\mu^{u-d}}{p}$.  The
axial charge $g_A$ and charge radius $r_A$ are defined at 
zero-momentum transfer $Q^2=0$: $G_A(0) \equiv g_A$ and $\langle
r_{A}^2\rangle = -6\frac{d}{dQ^2}
\left.\left(\frac{G_{A}(Q^2)}{G_{A}(0)}\right)\right|_{Q^2=0}$.

\begin{figure}[!tbh]
  \centering
  \includegraphics[width=0.32\textwidth]{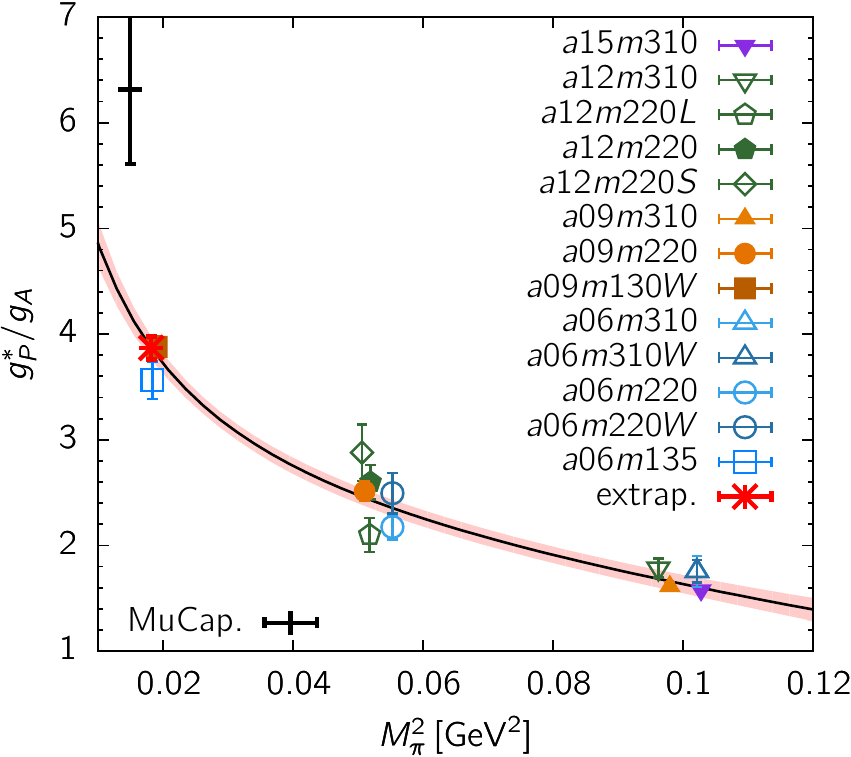}
  \includegraphics[width=0.32\textwidth]{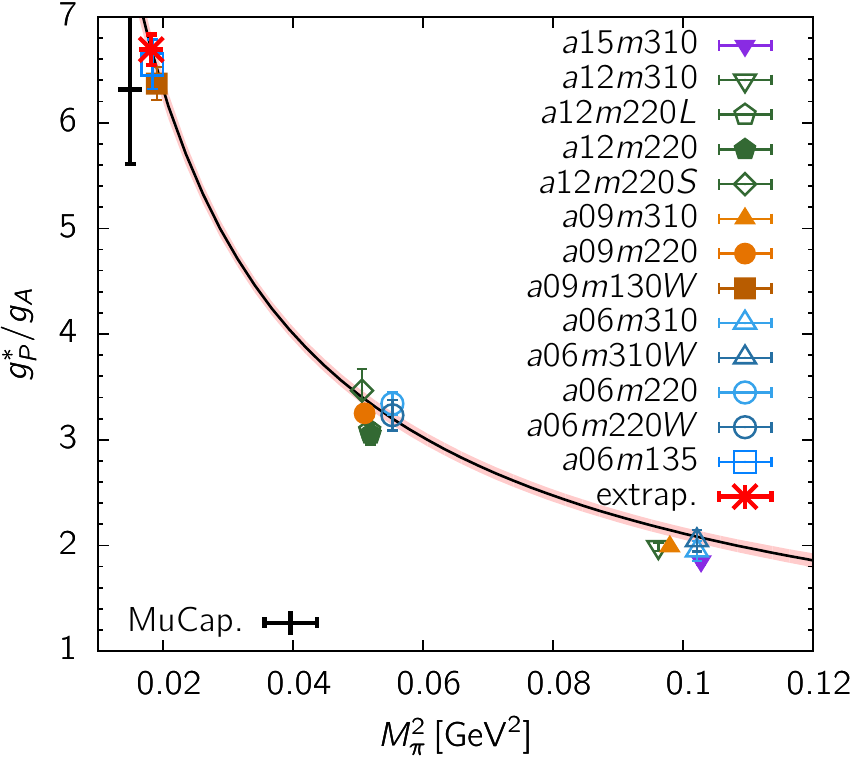} 
  \includegraphics[width=0.32\textwidth]{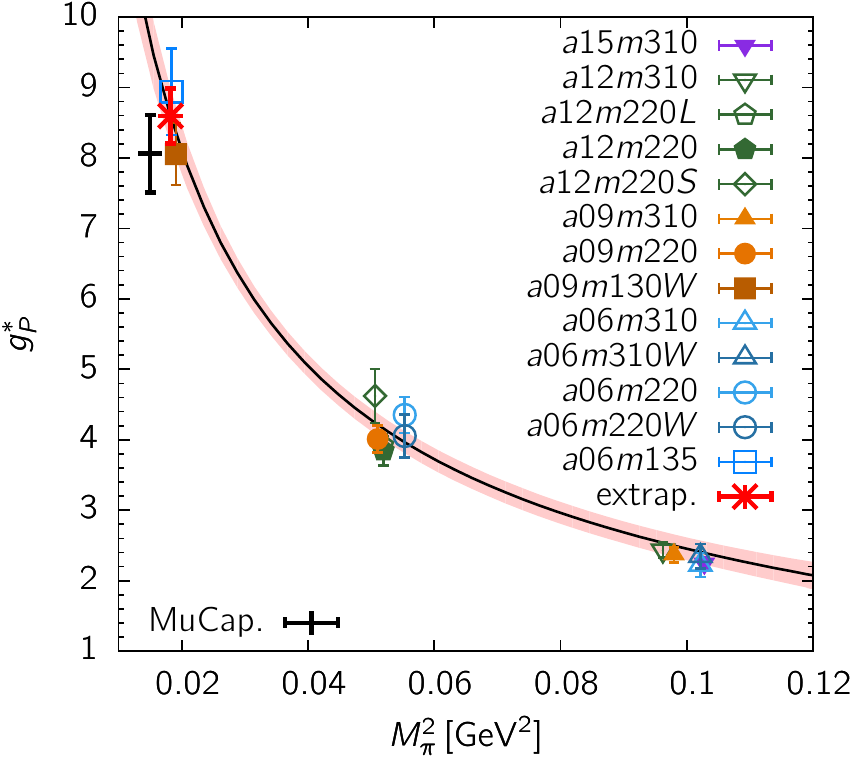}
  \caption{Chrial continuum fit using Eq.~\protect\eqref{eq:gPstar-CC-extrap} to $g_P^\ast/g_A =
    g_P^{\ast\;\text{(bare)}}/g_A^\text{(bare)}$ data with 
    $S_\text{2pt}$ (left) and $S_{A4}$ with $g_A$ from the
    $z^2$-fit to $G_A(Q^2\neq0)$ data (middle).  (right) $g_P^\ast =
    Z_A g_P^{\ast\;\text{(bare)}}$, where $g_P^{\ast\;\text{(bare)}}$ data are the
    same as in the middle panel and $Z_A$ is the axial current
    renormalization factor that is calculated
    independently~\cite{Gupta:2018qil}.\looseness-1}
  \label{fig:gPstar}
\end{figure}

The lattice axial form factor $G_A$ has traditionally been extracted
by fitting the spectral decomposition of the 3-point functions using
the spectrum taken from fits to 2-point functions (called the standard
strategy $S_\text{2pt}$ hereafter). These data 
have shown significant violation of the PCAC relation, 
\begin{align}
  \text{PCAC}:\; 2 {\widehat m} G_P(Q^2) &= 
  2 M G_A(Q^2) - \frac{Q^2}{2M} {\widetilde G}_P(Q^2) \,,
  \label{eq:PCAC}
  \quad
  \text{PPD}:\; \widetilde{G}_P(Q^2) = \frac{4M^2}{Q^2+M_\pi^2} G_A(Q^2) \,,
\end{align}
among the three form factors~\cite{Rajan:2017lxk}, as shown in
the top left panel of Fig.~\ref{fig:PPD-PCAC}. Here $\hat{m}$ is the PCAC
quark mass. Analogously, the induced pseudoscalar form factor
$\widetilde{G}_P$ with $S_\text{2pt}$ does not follow the $M_\pi^2$
behavior predicted by pion-pole dominance (PPD)
(Fig.~\ref{fig:PPD-PCAC} bottom left panel), and gives a smaller
$g_P^\ast \sim 0.6 g_P^\ast\vert_\text{exp}$ than the experimental
value (Fig.~\ref{fig:gPstar} left panel).
Fits to the axial form factor give a much larger axial mass than
obtained from a phenomenological parameterization of neutrino-nucleus
scattering data as shown in Fig.~\ref{fig:GA-comp-S2pt-SA4} left
panel. These deviations in $G_A$ result in a $g_A$ that is smaller than
the experimental value $g_A = 1.2766(20)$, and also a smaller charge radius
$r_A$.

In Ref.~\cite{Jang:2019vkm}, we showed that $S_\text{2pt}$ does not
include a lower energy excited state that dominates the 3-point
correlator with the insertion of $A_4$. This correlator,
$C_{A4}^\text{3pt}$, had been neglected in previous analyses because
of the large ESC and the failure of $S_\text{2pt}$ to fit it. We show that the 
large contribution of excited-states to $C_{A4}^\text{3pt}$ 
allows us to isolate a ``state'' with energy 
close to the noninteracting $N(\vec{0})\pi(\vec{p})$ state (or 
$N(-\vec{p})\pi(\vec{p})$ depending on the momentum
combinations). The new strategy, $S_{A4}$ with excited state energy
from the $C_{A4}^\text{3pt}$ and ground state mass $M$ and energy
$E(\vec{p})$ from the two-point correlators, fixes PCAC and gives
$g_P^\ast \approx g_P^\ast\vert_\text{exp}$. This was demonstrated
using the physical pion mass ensemble $a09m130W$ in
Ref.~\cite{Jang:2019vkm} and discussed in Section~\ref{sec:PCAC}.

In the following sections, we present a comparison of the axial form
factors extracted from the two strategies, $S_\text{2pt}$ and
$S_{A4}$~\cite{Jang:2019vkm} for 13 calculations on 11 ensembles of
$2+1+1$-flavor HISQ lattice generated by the MILC
collaboration~\cite{Bazavov:2012xda}. Details of the lattice
parameters, statistics accumulated and the spectrum used in the strategy
$S_\text{2pt}$ can be found in Ref.~\cite{Jang:2019jkn}.~\footnote{The
  physical mass ensemble $a06m135$ now has an additional 399
  configurations.}

\section{PCAC and Pion-pole Dominance (PPD)}
\label{sec:PCAC}

\begin{figure}[!tbh]
  \centering
  \includegraphics[width=0.43\textwidth]{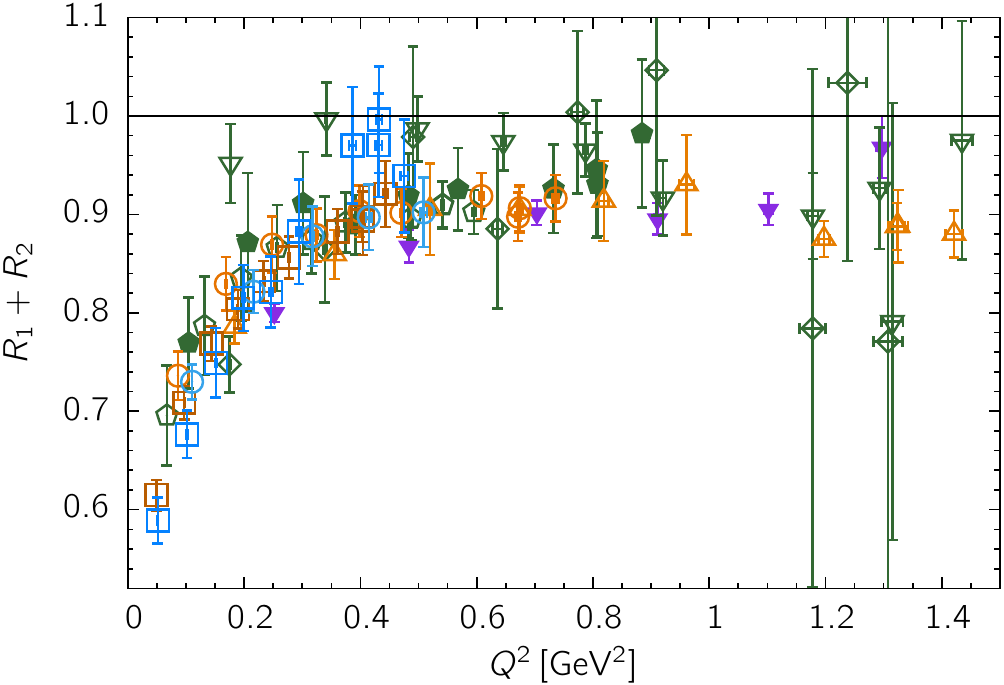} 
  \hspace{1em}
  \includegraphics[width=0.43\textwidth]{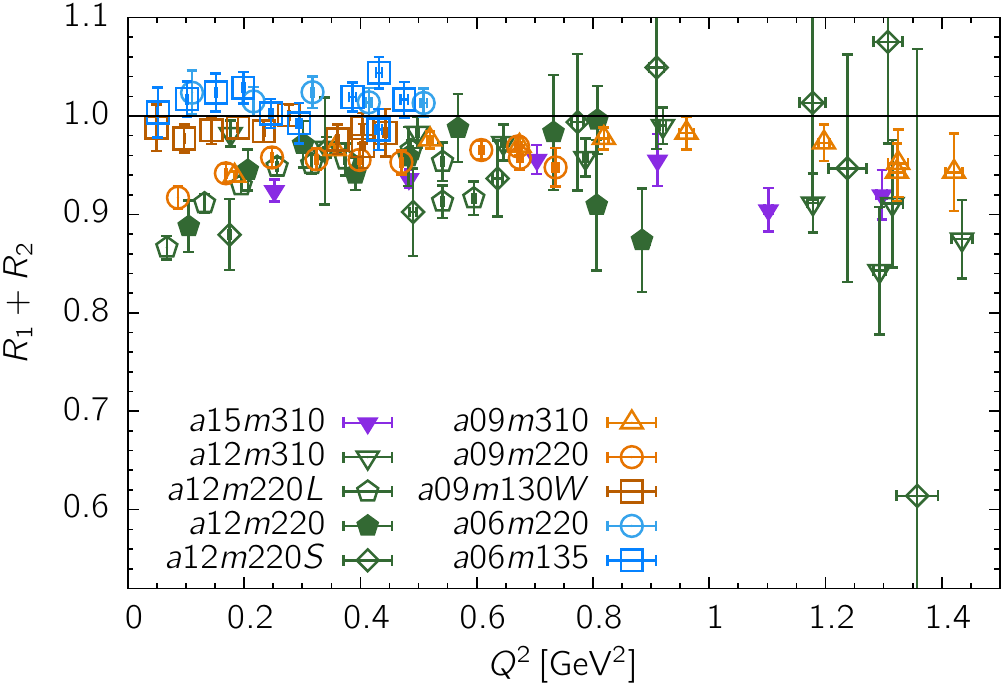}
  \\ \vspace{0.2em}
  \includegraphics[width=0.43\textwidth]{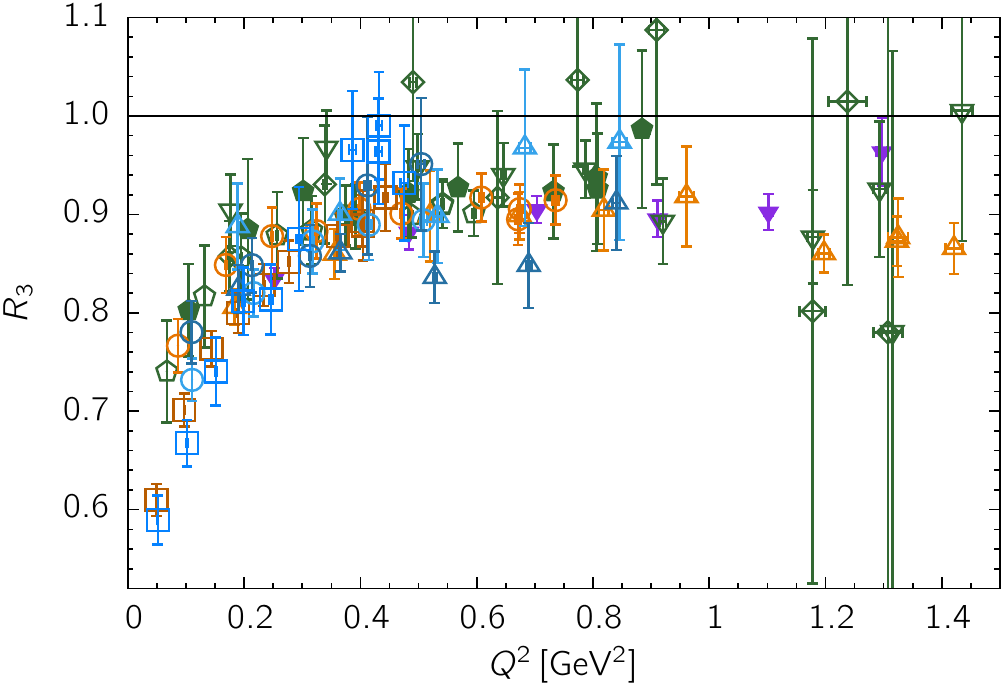} 
  \hspace{1em}
  \includegraphics[width=0.43\textwidth]{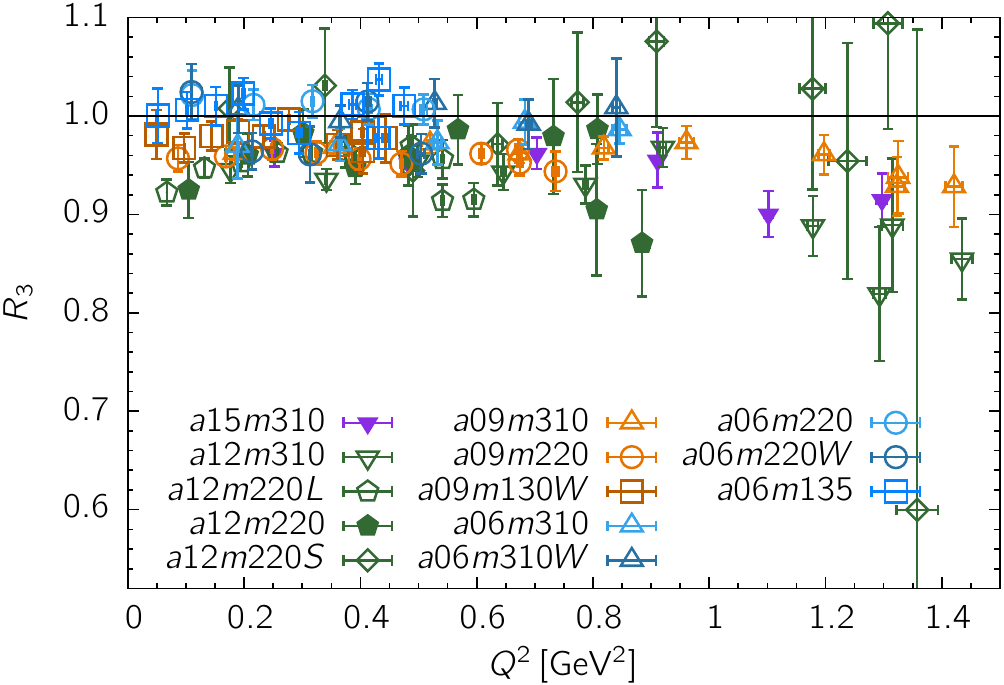} 
  \caption{Tests of the PCAC relation $R_1+R_2=1$ and the PPD $R_3=1$ for (left) $S_\text{2pt}$ and (right) $S_{A4}$.}
  \label{fig:PPD-PCAC}
\end{figure}

Tests of the PCAC relation and PPD hypothesis given in
Eq.~\eqref{eq:PCAC} are shown in Fig.~\ref{fig:PPD-PCAC} in term of the ratios $R_1+R_2$ and
$R_3$ with 
\begin{align}
  R_1 =&\;  \frac{Q^2}{4 M^2} \frac{{\tilde G}_P(Q^2)}{G_A(Q^2)}\,,\quad
  R_2 = \frac{2 {\widehat m}}{2M} \frac{G_P(Q^2)}{G_A(Q^2)} \,,\quad
  R_3 = \frac{Q^2+M_\pi^2}{4 M^2} \frac{{\tilde G}_P(Q^2)}{G_A(Q^2)} \,.
\end{align}
The left panels 
show increasing violation of PCAC and PPD with $S_\text{2pt}$ as $Q^2
\to0$, $M_\pi \to 135\,\MeV$ and $a \to 0$, ie, deviation from unity. This violation is reduced
to $\lesssim 5\%$ with $S_{A4}$ as shown in the right panels. Once
PPD is demonstrated, the improvement in the continuum limit result for
the induced pseudoscalar coupling $g_P^\ast$ follows as a consequence as discussed in Sec.~\ref{sec:GP}.

\section{Axial Form Factor $G_A$}
\label{sec:aff}

\begin{figure}[!tbh]
  \centering
  \includegraphics[width=0.495\textwidth]{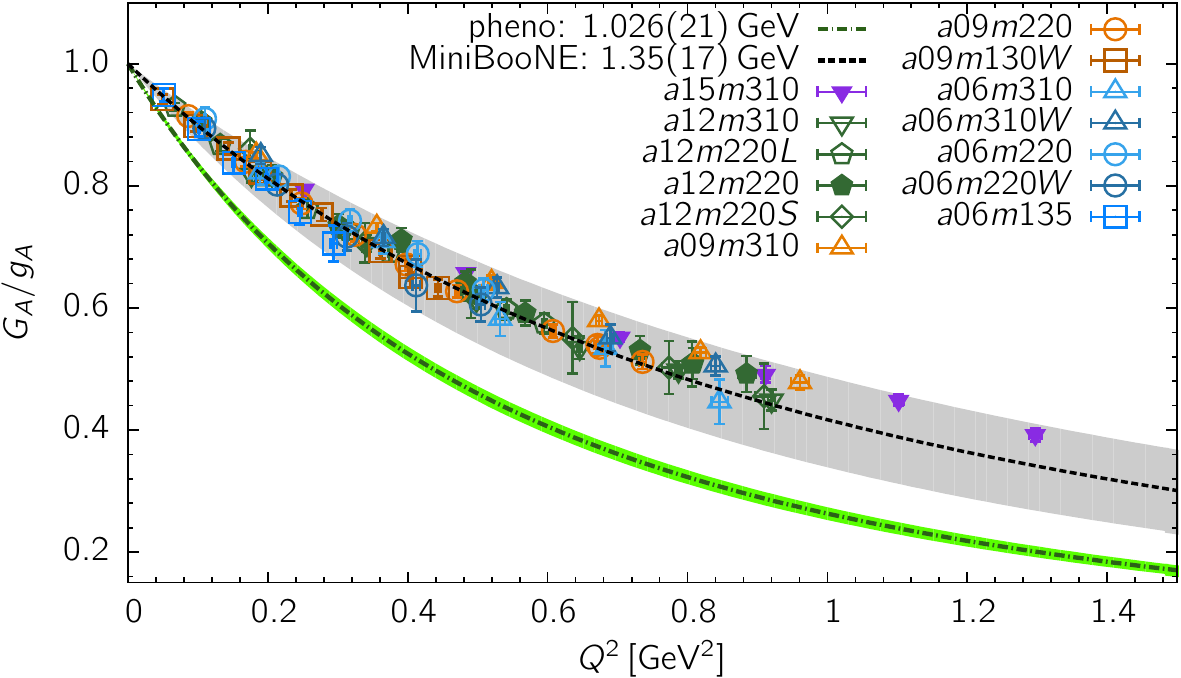} \hfill
  \includegraphics[width=0.495\textwidth]{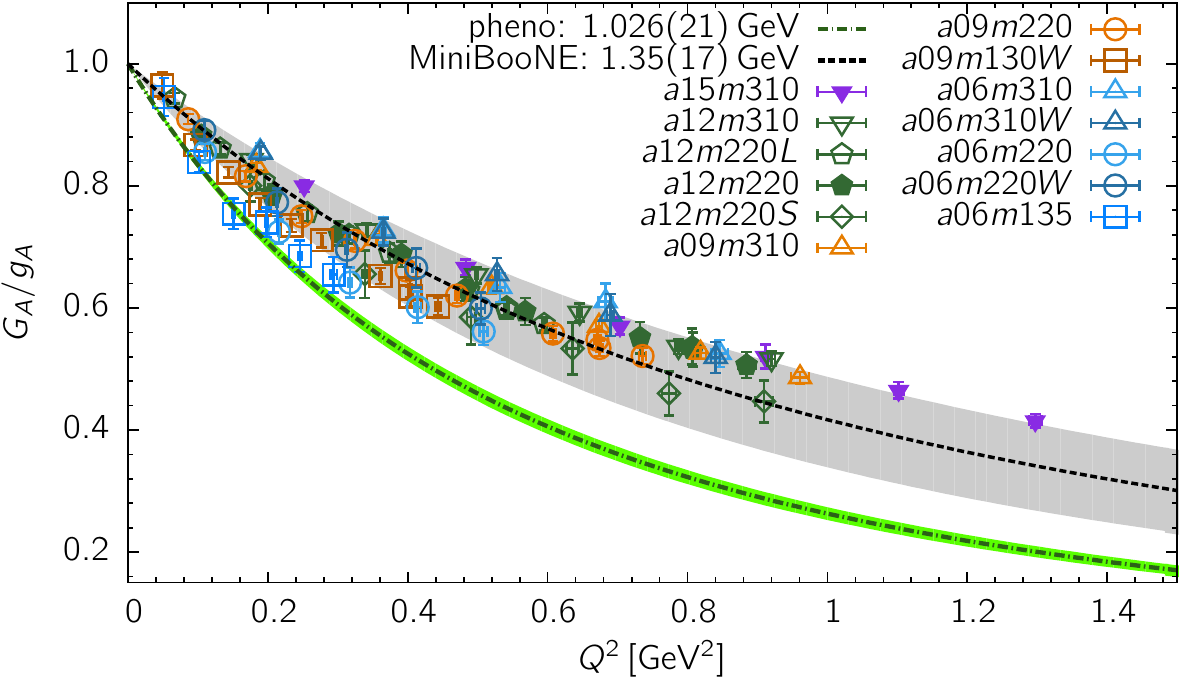}
  \caption{Comparing the axial form factor $G_A$ with charge $g_A$
    extracted using strategy $S_\text{2pt}$ (left) with $G_A$ from 
    strategy $S_{A4}$ and normalized by $g_A$ obtained from a $z^2$-fit ($z^1$-fit for the two physical ensemble) to it.}
  \label{fig:GA-comp-S2pt-SA4}
\end{figure}

In Fig.~\ref{fig:GA-comp-S2pt-SA4}, the axial form factor $G_A$ from
two strategies $S_\text{2pt}$ and $S_{A4}$ are compared.  The lattice
data for $G_A$ from $S_\text{2pt}$ are systematically above the
phenomenological curve (dipole fit with $M_A=1.026(21)$~GeV) for the
available kinematic range $0 < Q^2 \lesssim 1.4\,\GeV^2$, while
variations in lattice spacing and pion mass are smaller than the
deviation from that curve.
Compared to Refs.~\cite{Jang:2018lup,Jang:2018djx,Rajan:2017lxk}, the
increased statistics, number of data sets ($8\to13$), and the number
of excited states included in the three-point correlator fits ($2\to
3^\ast$) do not result in a significant change in the data or the
overall picture, and the deviation persists.  The spread with the
strategy $S_{A4}$ is larger as shown in
Fig.~\ref{fig:GA-comp-S2pt-SA4} (right).  Note that different $g_A$
are used to normalize $G_A(Q^2\neq0)$ in the two panels in
Fig.~\ref{fig:GA-comp-S2pt-SA4}: $g_A$ from a direct fit to
$C_{A3}^\text{3pt}(\vec{p}=\vec{0})$ when using $S_\text{2pt}$, and
extrapolation of the z-expansion fit to the nonzero momentum transfer
data obtained with $S_{A4}$.

While one can extract $g_A$ from a direct fit to
$C_{A3}^\text{3pt}(\vec{p}=\vec{0})$ with $S_\text{2pt}$ but, because the $A_4$ correlator vanishes at zero momentum, 
there is no information on the relevant excited states with $S_{A4}$ from the $A_4$ 
channel.  Thus, within $S_{A4}$ it is not obvious what value of $g_A$
to use to plot $G_A(Q^2)/g_A$. One can determine $g_A$ by
extrapolating the $G_A(Q^2\neq 0)$ data using the z-expansion or the
dipole ansatz.  We find that these estimates have a larger uncertainty
and the dipole ansatz does not fit the small $Q^2$ data with $S_{A4}$ in most
cases. One can also extract it by assuming that $N(\bm{p}=1)
\pi(\bm{p}=-1)$ is the relevant lightest excited state or by leaving
the first excited state energy a free parameter in the fits used to
remove ESC.  Note that when using these alternate methods for $g_A$ 
with $S_{A4}$, the renormalized form factor $G_A/g_A$ can
[under]overshoot the expected value 1.0 at $Q^2 = 0$. Also, the
difference in $G_A(Q^2)$ between $S_\text{2pt}$ and $S_{A4}$ is mainly
manifest at the smallest $Q^2$ on the physical mass ensembles.  Thus,
it is important to control this systematic, and we are still exploring 
options to get reliable estimates and defensible uncertainty quantification in them.

\begin{figure}[!tbh]
  \centering
  \includegraphics[width=0.342\textwidth]{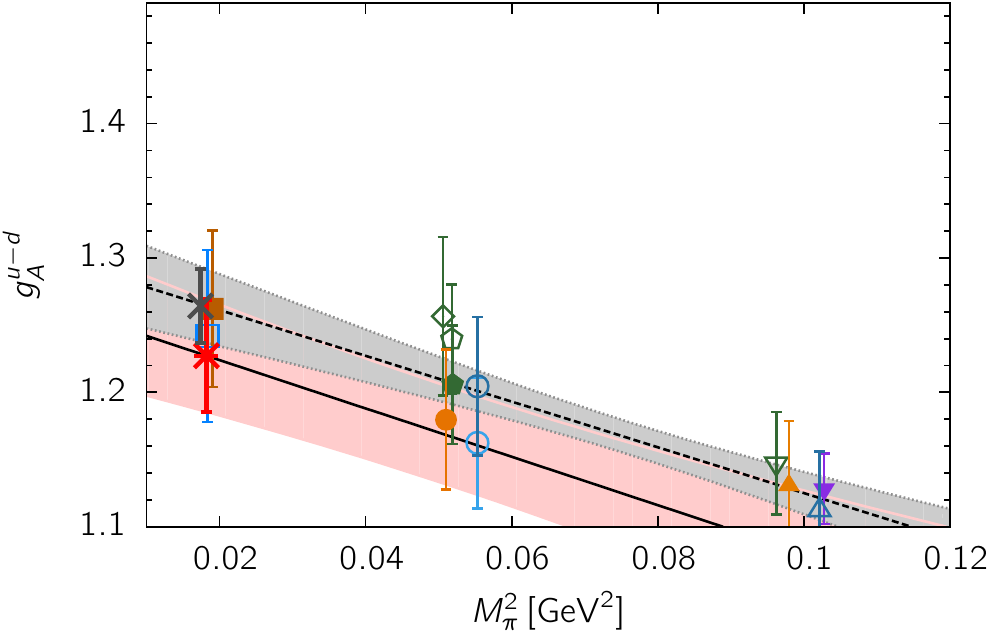} 
  \includegraphics[trim=20 0 0 0,clip,width=0.32\textwidth]{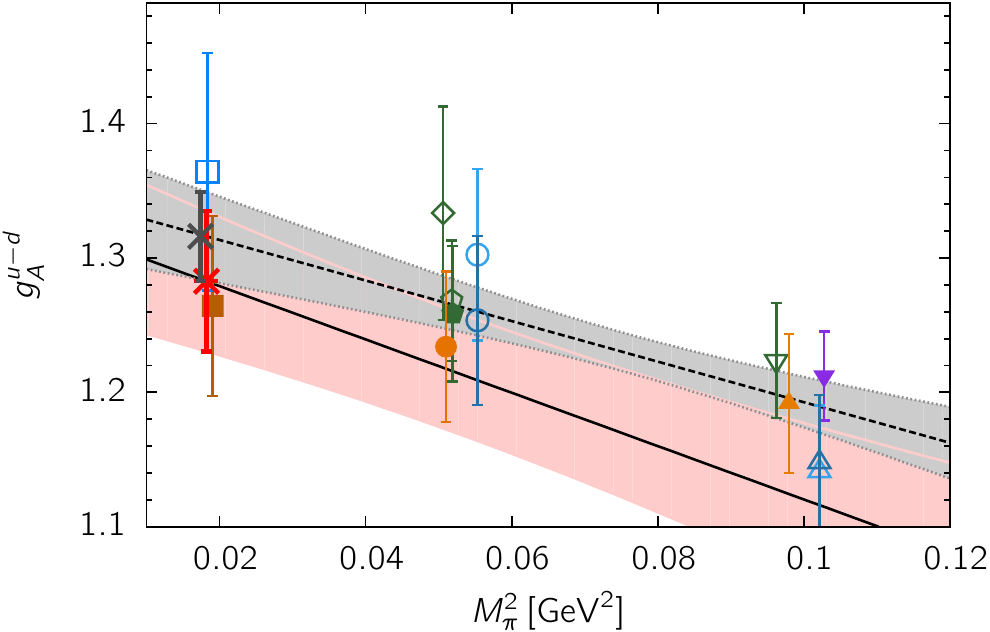} 
  \includegraphics[trim=20 0 0 0,clip,width=0.32\textwidth]{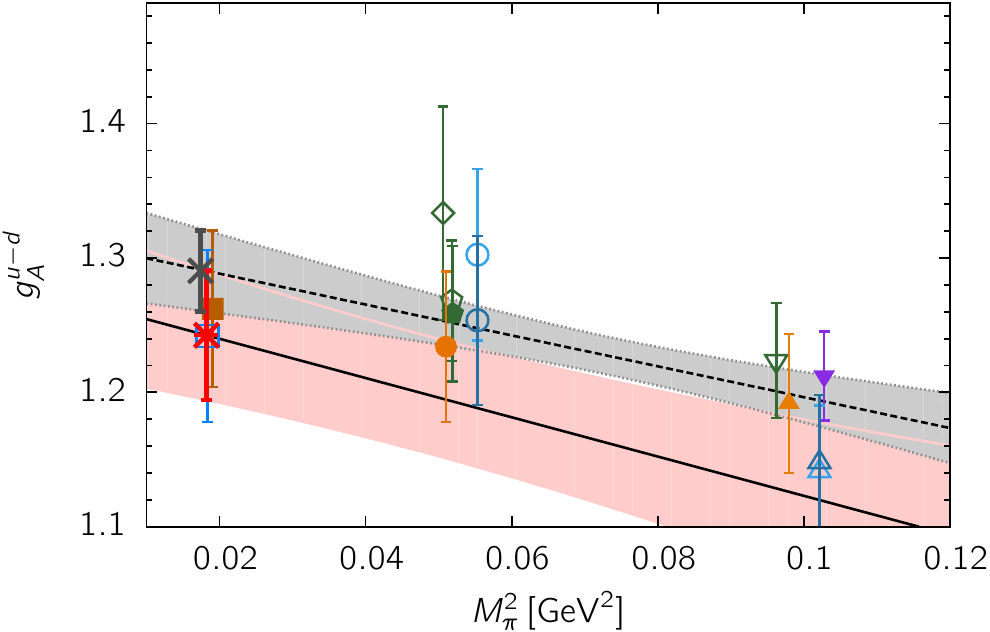} \\ \vspace{0.2em}
  \includegraphics[width=0.342\textwidth]{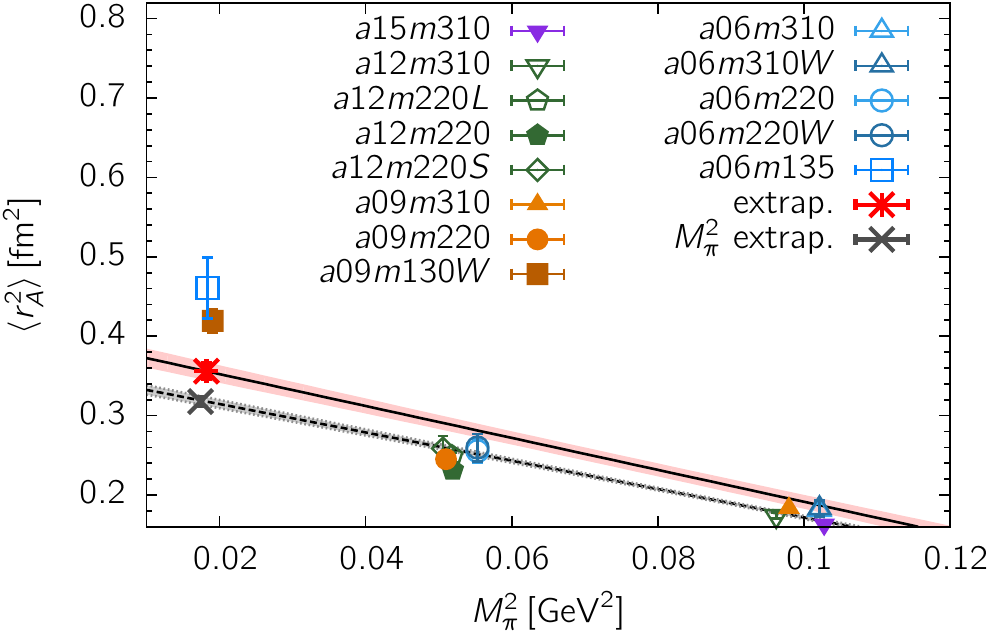} 
  \includegraphics[trim=20 0 0 0,clip,width=0.32\textwidth]{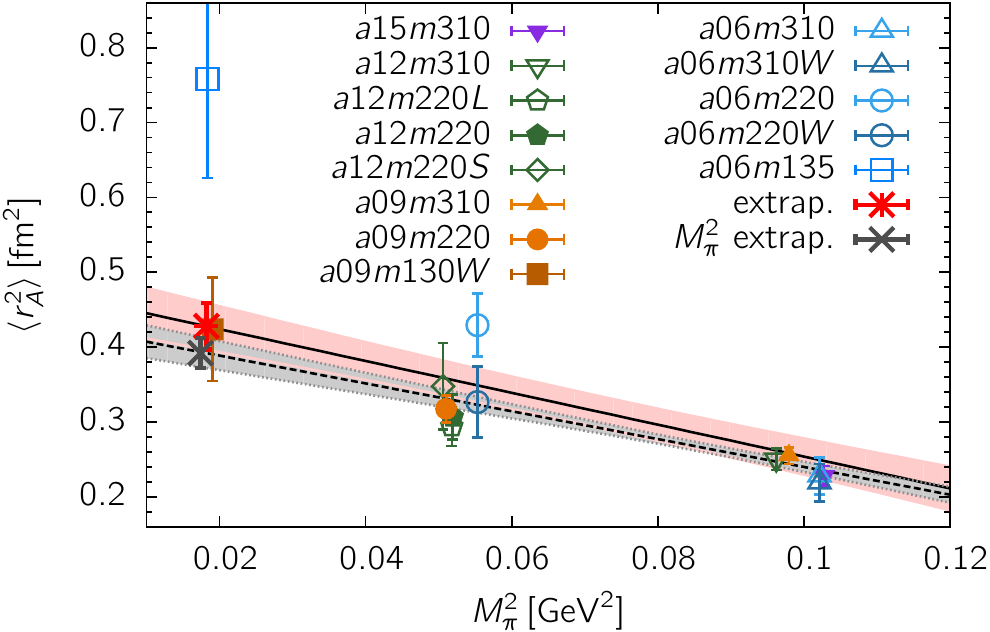}
  \includegraphics[trim=20 0 0 0,clip,width=0.32\textwidth]{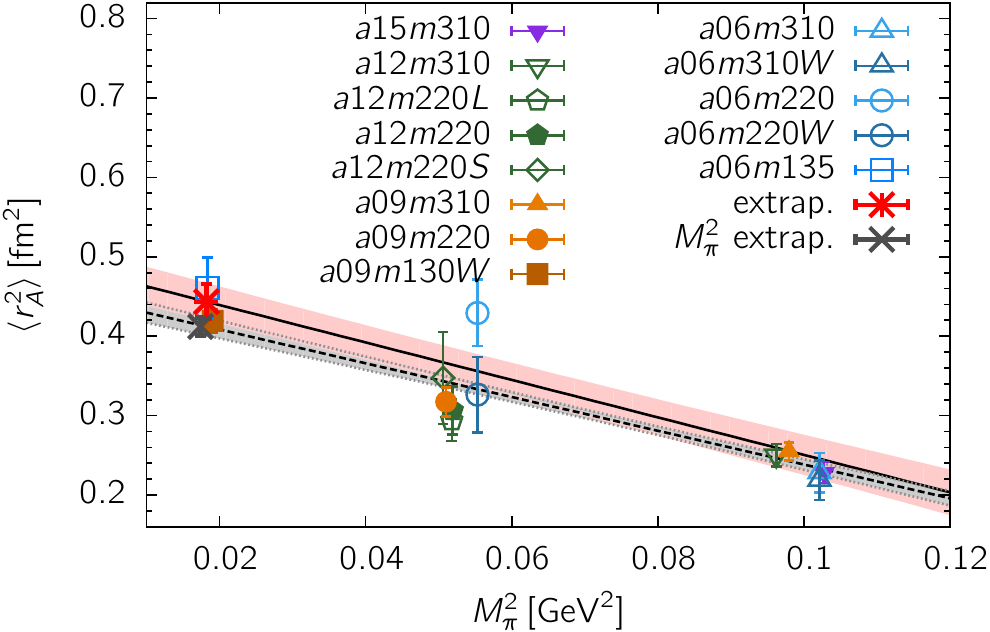}
  \caption{Data and CCFV fits (the pink band with the black solid
    line) using the ansatz $y=b_0 + b_1 a + b_2 M_\pi^2 + b_3 M_\pi^2
    \exp(-M_\pi L)$ for $g_A$ (top) and $r_A^2$ (bottom). On each
    ensemble, $ g_A$ and $r_A^2$ are obtained using the $z$-expansion
    with order $k=1$ (left), $k=2$ (middle), and $k=1$ for the two
    physical mass ensembles and $k=2$ for the rest (right). The grey
    band with the black dashed line is a fit with $b_1=b_3=0$. \cut{The
    data symbols used are the same as labeled in
    Fig.~\protect\ref{fig:GA-comp-S2pt-SA4}. }Result of the
    extrapolation are given in Table~\ref{tab:CCFV-gA-rA}.}
  \label{fig:CCFV-gA-rA}
\end{figure}

Preliminary results for the axial charge $g_A$ and $r_A^2$ obtained
using the $z$-expansion fits are shown in Fig.~\ref{fig:CCFV-gA-rA}
along with the continuum-chiral-finite-volume (CCFV) fits to get their
values at $a\to 0$, $M_\pi L \to \infty$ and $M_\pi = 135\,\MeV$.  In
the fits, we imposed the bound $\abs{a_i} \leq 5$ on the $z$-expansion
coefficients by using gaussian priors to stabilize higher order fits
and applied a cutoff $Q^2_\text{cut} \sim 1$~GeV, above which the
extraction of the data are not yet reliable. Overall, the
$z^k$-expansion fits with order $k=1$ and $k=2$ describe the data
well, and the changes in $g_A$ and $r_A^2$ with $k \ge 2$ are
small. Results from the CCFV fits are given in
Table~\ref{tab:CCFV-gA-rA}.  At present, the crucial $a06m135$
physical mass ensemble data have large errors, so we are working to
increase the statistics and thereby improve the reliability of the
CCFV fits.

\begin{table}[!tbh]
  \centering
  {\small%
  \begin{tabular}{l|ccc|ccc}
    \hline\hline
    Ref. & $g_A$ &  $\chi^2/\text{dof}$ & $p$-value & $\expv{r_A^2}\,[\fm^2]$ &  $\chi^2/\text{dof}$ & $p$-value \\
    \hline
    Fig.~\ref{fig:CCFV-gA-rA}, left   & 1.23(4) & 0.19 & 0.99 & 0.356(10) & 7.64 & 3$\times 10^{-11}$ \\
    Fig.~\ref{fig:CCFV-gA-rA}, middle & 1.28(5) & 0.34 & 0.96 & 0.428(31) & 1.90 & 0.05 \\
    Fig.~\ref{fig:CCFV-gA-rA}, right  & 1.24(5) & 0.37 & 0.95 & 0.444(22) & 1.34 & 0.21 \\
    \hline\hline
  \end{tabular}}
  \caption{Results for $g_A$ and $r_A^2$ from the CCFV fits shown in Fig.~\protect\ref{fig:CCFV-gA-rA}.}
  \label{tab:CCFV-gA-rA}
\end{table}

\section{Induced Pseudoscalar $\widetilde{G}_P$ and Pseudoscalar $G_P$ Form Factors}
\label{sec:GP}

\begin{figure}[!tbh]
  \centering
  \includegraphics[width=0.495\textwidth]{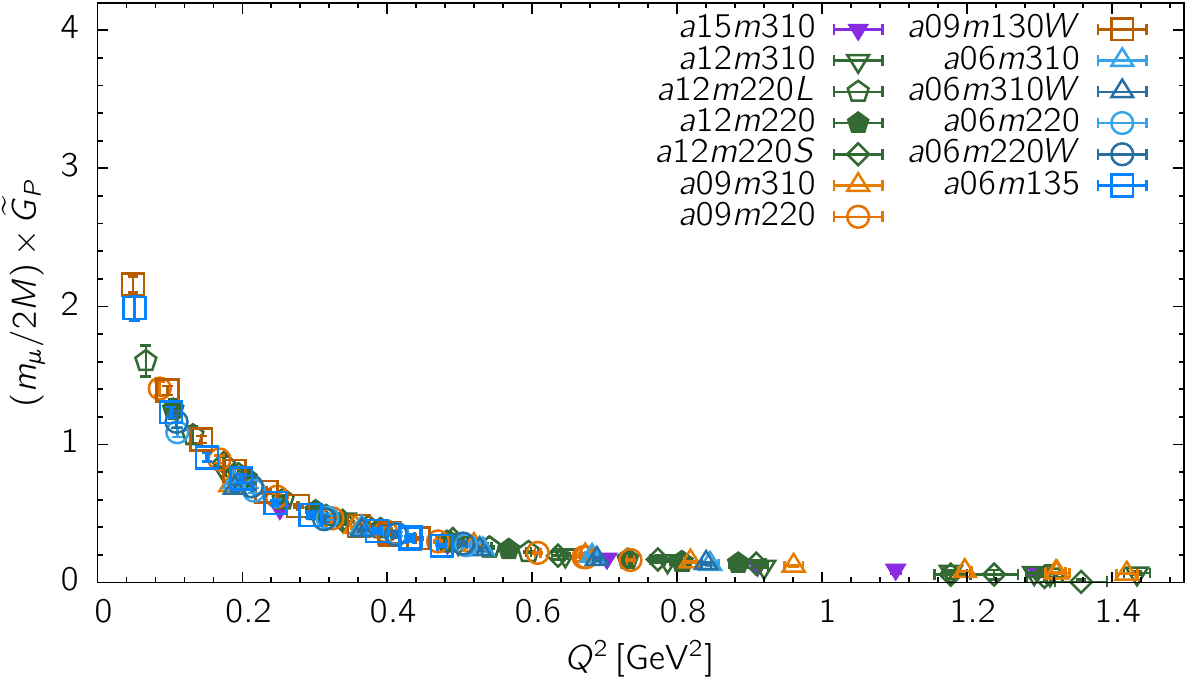} \hfill
  \includegraphics[width=0.495\textwidth]{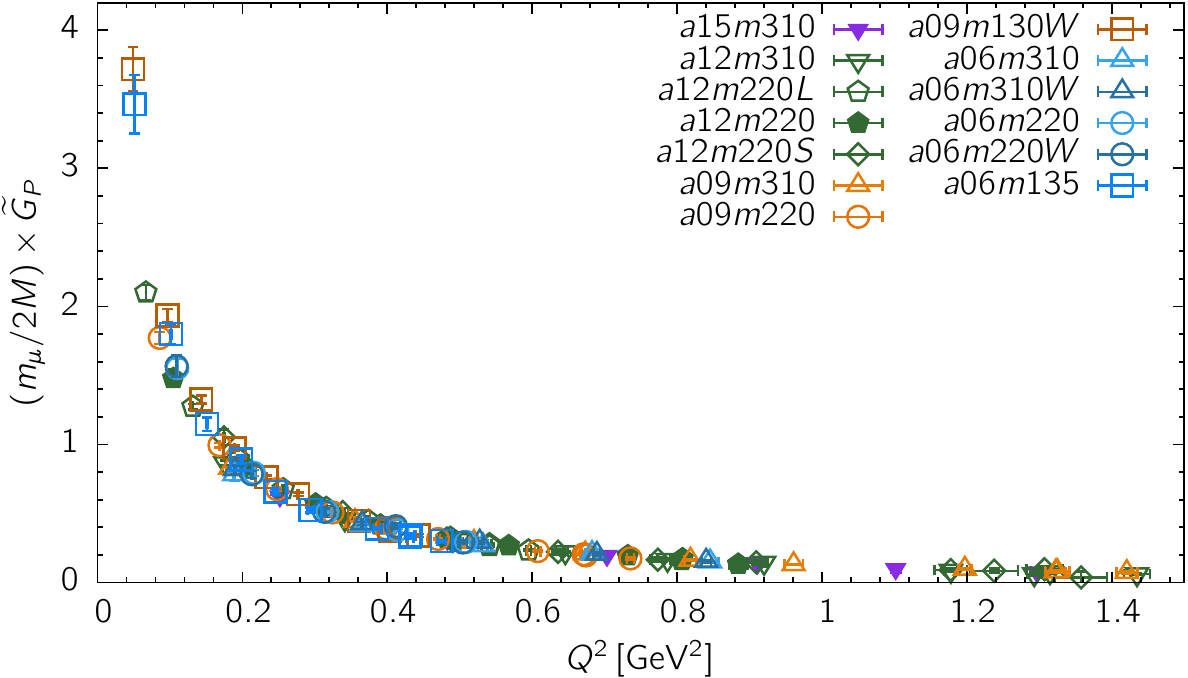}
  \\ \vspace{0.2em}
  \includegraphics[width=0.495\textwidth]{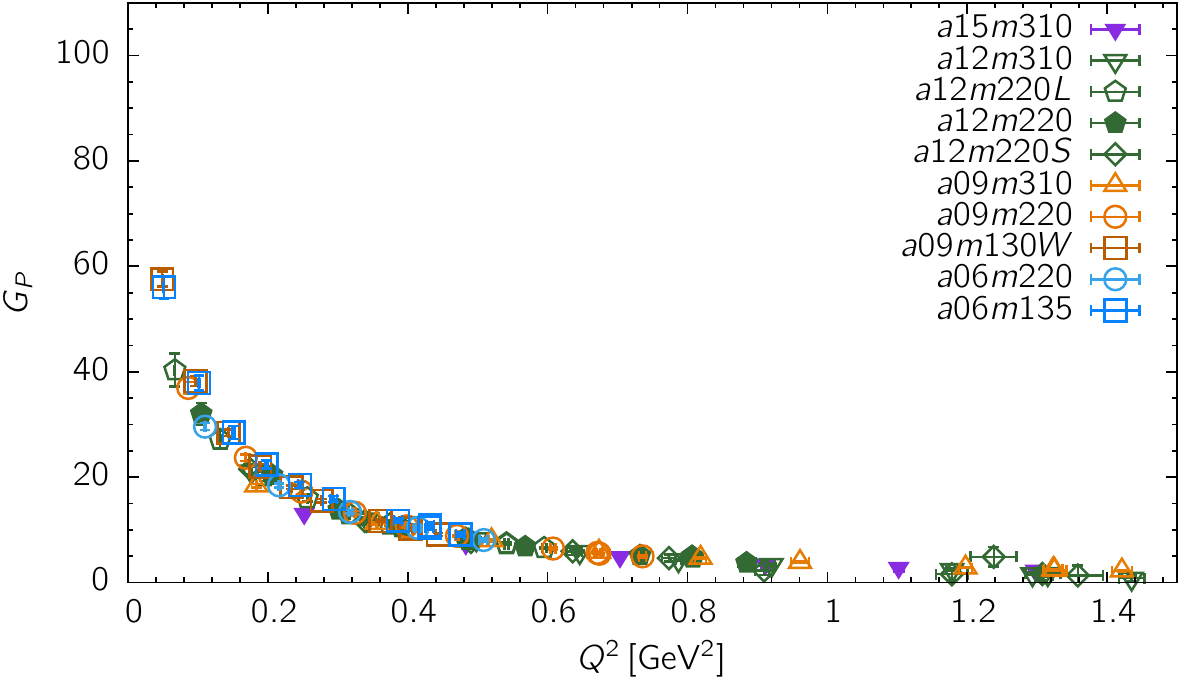} \hfill
  \includegraphics[width=0.495\textwidth]{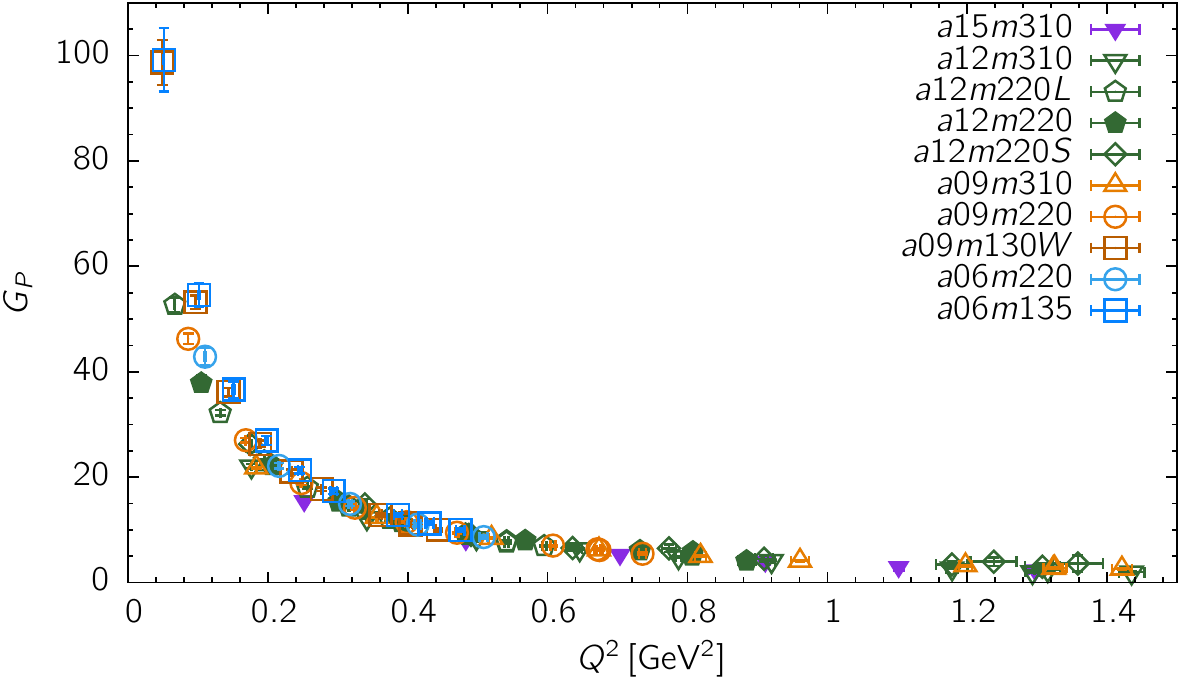} 
  \caption{Comparison of $(m_\mu/2M)\times\widetilde{G}_P$ (top) and
    $G_P$ (bottom) obtained using the two strategies $S_\text{2pt}$
    (left) and $S_{A4}$ (right). Here, the form factors are not
    renormalized.}
  \label{fig:GP-GPS}
\end{figure}

Data for the induced pseudoscalar $\widetilde{G}_P$ and pseudoscalar
$G_P$ form factors from two strategies $S_\text{2pt}$ and $S_{A4}$ are
summarized in Fig.~\ref{fig:GP-GPS}. Both form factors, and
consequently the induced pseudoscalar coupling $g_P^\ast \equiv
(m_\mu/2M)\widetilde{G}_P(0.88m_\mu^3)$, are clearly
enhanced at the small $Q^2$ with $S_{A4}$. Note that we do not have data
for $G_P$ on the three ensembles $a06m310$, $a06m310W$, $a06m220W$.

To extract $g_P^\ast$ on each ensemble, we fit $\widetilde{G}_P$ using 
Eq.~\eqref{eq:gPstar-kin-extrap} and read off the value at the 
kinematic point $Q^2=0.88m_\mu^2$. Eq.~\eqref{eq:gPstar-kin-extrap} includes 
the pion-pole and the leading analytical behavior. 
\begin{align}
  \tilde{G}_P(Q^2) =&\; \frac{c_0}{Q^2+M_\pi^2} + c_1 + c_2 Q^2 \,.
  \label{eq:gPstar-kin-extrap} \\
  Y_{1,2}(a,M_\pi)\cut{_{1,2}} =&\; \frac{d_1}{M_\pi^2 + 0.88m_\mu^2} + d_2 + d_3 a + d_4 M_\pi^2 \,.
  \label{eq:gPstar-CC-extrap}
\end{align}
Result at the physical point, $a\to 0$ and
$M_\pi=135\,\MeV$, is obtained using Eq.~\eqref{eq:gPstar-CC-extrap} with the 
data renormalized using two different methods: $Y_1 = g_P^\ast/g_A =
g_P^{\ast\;\text{(bare)}}/g_A^\text{(bare)}$ and $Y_2 = g_P^\ast = Z_A
g_P^{\ast\;\text{(bare)}}$.
The $g_P^\ast$ with 
strategy $S_{A4}$ is consistent with the experimental value
$g_P^\ast\vert_\text{MuCap} = 8.06(55)$~\cite{Andreev:2012fj,Andreev:2015evt} obtained from the
MuCap experiment $[\mu^- + p \to \nu_\mu + n]$, or
$g_P^\ast/g_A\vert_\text{MuCap} = 6.31(70)$ with
$g_A\vert_\text{exp}=1.2766(20)$. \looseness-1

\begin{table}[!tbh]
  \centering
  {\small%
  \begin{tabular}{lccccccc}
    \hline\hline
    Ref. & $g_P^\ast/g_A$ & $d_1\;[\GeV^{2}]$ &  $d_2$ & $d_3\;[\fm^{-1}]$ & $d_4\;[\GeV^{-2}]$ & $\chi^2/\text{dof}$ & $p$-value \\
    \hline
    Fig.~\ref{fig:gPstar}, left  & 3.87(11) & 0.056(09) & 1.91(31) & -0.08(70)   & -8.0(2.5) & 2.27 & 0.02 \\
    Fig.~\ref{fig:gPstar}, middle & 6.69(14) & 0.159(09) & 0.78(27) & -1.66(55)   & -1.4(1.8) & 1.27 & 0.24 \\
    \hline\hline 
    Ref. & $g_P^\ast$ & $d_1\;[\GeV^{2}]$ &  $d_2$ & $d_3\;[\fm^{-1}]$ & $d_4\;[\GeV^{-2}]$ & $\chi^2/\text{dof}$ & $p$-value \\
    \hline
    Fig.~\ref{fig:gPstar}, right  & 8.60(39)    & 0.208(26) & 0.94(73) & -1.11(1.33) & -4.0(5.1) & 0.91 & 0.51 \\
    \hline\hline
  \end{tabular}}
  \caption{Results for the chiral-continuum extrapolation of $g_P^\ast/g_A$ and $g_P^\ast$ using Eq.~\eqref{eq:gPstar-CC-extrap}.}
  \label{tab:gPstar-CC-extrap}
\end{table}

\section{Summary}
\label{sec:last}

We present a comparison between two strategies, $S_\text{2pt}$ and $S_{A4}$, 
for extracting the nucleon isovector axial form factors, $G_A$,
induced pseudoscalar $\widetilde{G}_P$, and pseudoscalar $G_P$.  The
new strategy $S_{A4}$ incorporates a low-energy state that was not
exposed by multistate fits to nucleon two-point correlators
constructed with a conventional nucleon interpolating operator.
Incorporating these lower lying excited states largely impacts the
pseudoscalar $G_P$ and induced pseudoscalar $\widetilde{G}_P$ form
factors. As a result, the PCAC relation and PPD hypothesis are now
resonably well satisfied: the deviation is reduced to about 10\% for
heavy pion mass ensembles and to only a few percent for the physical
mass ensembles.  Consequently, the induced pseudoscalar coupling
$g_P^\ast$ is also consistent with the experimental value.

The axial form factor $G_A$ shows a much smaller shift compared to $G_P$ and
$\widetilde{G}_P$ that is evident only at the smallest $Q^2$. Unfortunately,
becasue of the kinematic constraint, one cannot extract $g_A$ (the
data point at $Q^2=0$) with strategy $S_{A4}$.
The small $Q^2$ behavior is also critical for determining the axial
charge radius $r_A$. At discussed in Sec.~\ref{sec:aff}, analysis of
the small $Q^2$ (and $Q^2=0$) behavior is under progress and the
statistics on the $a06m135$ ensemble are being increased.

The new strategy $S_{A4}$ uses a 2-state fit for the three-point
correlator analysis. While inclusion of a single ``effective'' lower
energy excited state dramatically improves PCAC and PPD, it is
essential to develop methods that will provide a more detailed and
refined picture of the many possible excited states that can
contribute. Including the full tower of states that provide
significant ESC and controlling this systematics is necessary for
precision calculations of the axial form factors.

\section*{Acknowledgement}
\label{sec:ack}
\vspace{-3mm}
We thank the MILC Collaboration for providing the 2+1+1-flavor HISQ lattices.
The calculations used the Chroma software suite~\cite{Edwards:2004sx}.
Simulations were carried out on computer facilities at  (i) the National Energy
Research Scientific Computing Center, a DOE Office of Science User Facility
supported under Contract No. DE-AC02-05CH11231; and, (ii) the Oak Ridge Leadership Computing
Facility supported by the Office
of Science of the DOE under Contract No.
DE-AC05-00OR22725; (iii) the USQCD Collaboration, which are funded by the
Office of Science of the U.S. Department of Energy, and (iv) Institutional
Computing at Los Alamos National Laboratory.  T. Bhattacharya and R. Gupta were
partly supported by the U.S. Department of Energy, Office of Science, Office of
High Energy Physics under Contract No.~DE-AC52-06NA25396.  T. Bhattacharya, R.
Gupta, Y.-C. Jang and B.Yoon were partly supported by the LANL LDRD program.
Y.-C. Jang is partly supported by U.S. Department of Energy under Contract No.
DE-SC0012704.

\bibliography{lattice2019}

\end{document}